# Nucleation of Dislocations in 3.9 nm Nanocrystals at High Pressure

**Authors:** Abhinav Parakh[1], Sangryun Lee[2], K. Anika Harkins[3], Mehrdad T. Kiani[1], David Doan[4], Martin Kunz[5], Andrew Doran[5], Lindsey A. Hanson[3], Seunghwa Ryu[2] and X. Wendy Gu[4]*

**Affiliations:**

[1]Materials Science and Engineering, Stanford University, Stanford, CA 94305, USA.

[2]Mechanical Engineering, KAIST, Yuseong-gu, Daejeon 34141, Republic of Korea.

[3] Chemistry, Trinity College, Hartford, CT 06106, USA.

[4]Mechanical Engineering, Stanford University, Stanford, CA 94305, USA.

[5]Advanced Light Source, Lawrence Berkeley National Lab, Berkeley 94720, USA.

*Corresponding author:

X. Wendy Gu

452 Escondido Mall, Room 227, Stanford University, Stanford, CA 94305

650-497-3189

xwgu@stanford.edu

**Abstract:**

As circuitry approaches single nanometer length scales, it is important to predict the stability of metals at these scales. The behavior of metals at larger scales can be predicted based on the behavior of dislocations, but it is unclear if dislocations can form and be sustained at single nanometer dimensions. Here, we report the formation of dislocations within individual 3.9 nm Au nanocrystals under nonhydrostatic pressure in a diamond anvil cell. We used a combination of x-ray diffraction, optical absorbance spectroscopy, and molecular dynamics simulation to characterize the defects that are formed, which were found to be surface-nucleated partial dislocations. These results indicate that dislocations are still active at single nanometer length scales and can lead to permanent plasticity.

**Main text:**

**Introduction**

Permanent plastic deformation occurs in crystalline metals that are subjected to large strains at room temperature. This is due to irreversible interactions between dislocations, and between dislocations and microstructural features such as grain and twin boundaries. Recently, reversible deformation from large strains has been observed in a number of metallic nanostructures, such as sub-10 nm Ag nanocrystals (*1*) and 3.9 nm Au nanocrystals (*2*). Rapid diffusion of atoms at free surfaces and stress-induced diffusion at the nanocrystal-indenter and nanocrystal-substrate interfaces have been proposed as mechanisms for this pseudoelastic behavior (*1*)(*3*). Others have found that dislocations are involved in the deformation of single nanometer-sized nanocrystals and can contribute to reversible plasticity. For instance, surface-nucleated dislocations and deformation twinning have been observed in sub-10 nm nanowires, and stacking faults tetrahedra have been observed in sub-20 nm nanowires under tension in both experiments and molecular dynamics (MD) simulations (*4–7*). Dislocations and surface diffusion may also act cooperatively. *In situ* transmission electron microscope (TEM) tension tests on ~20 nm Ag nanowires showed that surface diffusion is enhanced at surface steps created by the passage of dislocations (*8*). Previous work from our group showed that pseudoelastic shape recovery in 3.9 nm Au nanocrystals is accompanied by the formation of irreversible defects, but the nature of the defects could not be determined (*2*).

These observations prompt the questions: Is there a limit to plasticity at small length scales? What is the smallest crystal in which dislocations can form and lead to irreversible deformation? This is critical to the processing and mechanical behavior of nanostructured materials such as nanocrystalline, nanotwinned and nanoporous metals, and the design of stable nano-devices with single nanometer metallic features (*9*). To answer these questions, deformation mechanisms in very small nanocrystals must be experimentally determined, but this remains challenging. *In situ* TEM mechanical testing is the leading method to investigate deformation mechanisms at this length scale, but results may be influenced by heating from the electron beam. In addition, fast dislocations and dislocations that are invisible at specific imaging conditions cannot be observed. X-ray diffraction (XRD) is another method to measure elastic strain and defect formation in metals under mechanical stress. The width and relative intensities of XRD peaks have previously been used to detect dislocation activity in nanocrystalline Ni under uniaxial tension (*10*) and compression in a diamond anvil cell (DAC) (*11*). These studies involve the response at grain boundaries as well as within the grains, so they cannot be directly applied to understand plasticity in individual nanocrystals. To do this, the structural response of isolated nanocrystals must be obtained. This presents a challenge for *in situ* XRD because the diffracted intensities from a single nanocrystal are much too small for detection.

Here, we use XRD to detect structural changes in an ensemble of monodisperse 3.9 nm Au nanocrystals that are compressed under a non-hydrostatic pressure in a DAC. Surfaces of the nanocrystals are protected by organic ligands, which prevents contact between the nanocrystals. Structural changes from XRD are corroborated with optical spectroscopy measurements, and MD simulations are used to determine the specific defects that correspond to the ensemble-averaged behavior from XRD. We show that irreversible deformation due to the formation of partial dislocations can occur in small metallic nanocrystals. This indicates that dislocation-mediated plasticity is still active at single nanometer length scales and must be considered in designing structures at this scale.

**Results and Discussion**

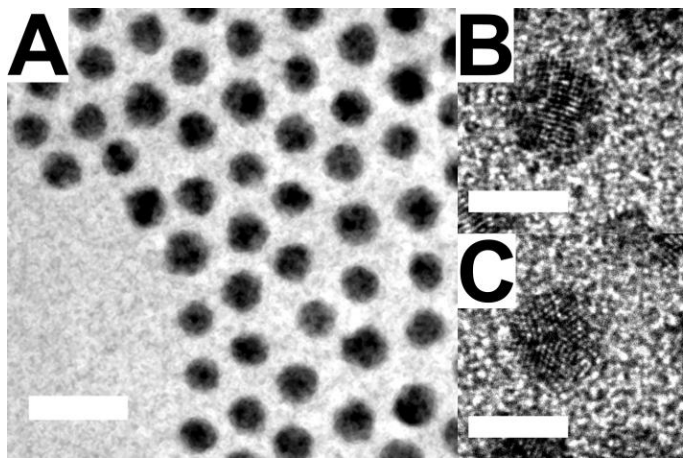

**Fig. 1. TEM images of nanocrystals.** A) Monodisperse 3.9 nm Au nanocrystals. Scale bar is 10 nm. High-resolution images of B) icosahedral and C) decahedral nanocrystals. Scale bar is 4 nm.

Au nanocrystals were synthesized using the organic phase reduction of chloroauric acid and capped with dodecanethiol ligands (*12*). The nanocrystal size distribution was found to be 3.9±0.6 nm using TEM (see Fig. 1A and Fig. S1). High-resolution TEM images showed that most of the identified nanocrystals were either icosahedral or decahedral in shape (Fig. 1B-C). Icosahedral nanocrystals have 20 twin boundaries, and decahedral nanocrystals have 5 twin boundaries. Ambient pressure XRD showed an FCC crystal structure, and significantly broader peaks than bulk Au due to the limited coherent scattering volume within the nanocrystals (Fig. S2). The (111), (220), (311) and (222) XRD peaks were shifted to higher 2θ angles by ~0.1º compared to that of the bulk, which corresponds to a ~1.8% volumetric compressive strain. This is due to compressive stress at the surface of the nanocrystal (*13*). The position of the (200) peak was shifted to lower 2θ angles by 0.15°. Broad shoulders were observed on the (200) and (220) peaks. These features are indicative of the high twin density in icosahedral and decahedral nanocrystals (*14*). The Debye scattering equation was used to fit the XRD pattern to determine the structure of the nanocrystals. In this method, the atomic positions for icosahedral and decahedral nanocrystals were generated for 1 to 6 nm diameter nanocrystals and used to simulate XRD patterns. A Rietveld-like refinement procedure was used to fit the experimental data (*14, 15*). The best fit was obtained by combining 60% icosahedral nanocrystals with a size distribution of 3.2±0.2 nm and 40% decahedral nanocrystals with a size distribution of 3.8±0.6 nm (Fig. S3). This result is in close agreement with the nanocrystal shape and size distribution observed in TEM.

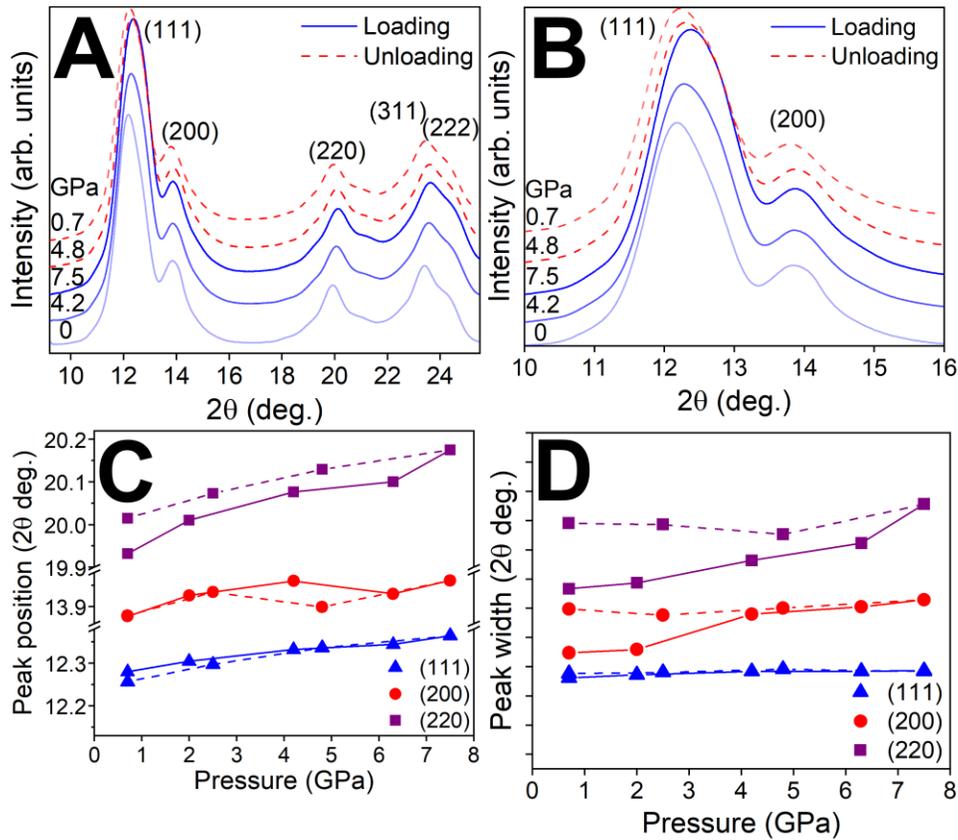

**Fig. 2. Experimental high-pressure XRD patterns.** A) All diffraction peaks and B) magnified view of (111) and (200) peaks. Change in diffraction peak C) position and D) width (each division is 0.1°), upon loading (solid line) and unloading (dashed line).

High pressure XRD was obtained during DAC compression experiments at the Advanced Light Source at Lawrence Berkeley National Laboratory (Fig. 2A-B). A non-hydrostatic pressure was applied to the nanocrystals by loading the nanocrystals as a thick film at the bottom of the DAC sample chamber, and using toluene as a non-hydrostatic pressure medium (*16*). XRD was collected while the nanocrystals were loaded up to 7.5 GPa and as pressure was released. The pressure was limited to 7.5 GPa to avoid sintering between the nanocrystals, which has been observed by our group and others at higher pressures (*17–19*). The XRD peak position and width (full width at half max) were observed to change with increasing and decreasing pressure and quantified at each pressure (Fig. 2C-D). The relative intensity of the XRD peaks does not change under pressure, which indicates that the nanocrystals remain randomly oriented.

The change in peak position indicates the elastic strain in the nanocrystals. The shift in the peak position shows that the lattice spacing decreases by 0.042 Å over 7.5 GPa and recovers to ~0.2% of its original value upon unloading. Due to the non-hydrostatic pressure, the change in lattice spacing is different along the loading axis (axial) and orthogonal to the loading axis (radial). The geometry of the X-ray setup is such that the measured lattice spacings correspond to planes that are almost aligned with the loading axis. Therefore, the measured change in lattice spacing is lower than in the hydrostatic case (Fig. S4). The difference between radial and axial stress components (termed as t) can give us an estimate of maximum deviatoric and shear stresses in the system. This difference can be calculated by considering the elastic anisotropy of a polycrystalline,

FCC metal. We used lattice strain theory to get a rough estimate of 't' (*20*, *21*) (see supplementary materials). Using this we estimated the maximum shear stress of Au nanoparticles to be about 2.3 GPa (Fig. S5).

Fig. 2D shows the change in peak width for the (111), (200) and (220) peaks with a complete pressure cycle. The (200) peak width showed a significant increase of 16% and the (220) peak width showed an increase of 23% with increasing pressure and remained at higher values after unloading. This indicates that irreversible deformation is occurring in the nanocrystals and remains in the nanocrystals on the time scale of the experimental measurements. The XRD peak width can be affected by changes in crystallite size, shape and microstrain (*22*). It is possible that crystalline domains within the nanocrystal become elongated under compression and split into smaller domains, but post-compression TEM images showed that the nanocrystal shape and size distribution is identical to that of the as-synthesized nanocrystals (Fig. S1). The (111) peak width is mostly affected by domain size changes and is least affected by the presence of defects like twinning and stacking faults in the nanocrystal (Fig. S6). The peak width for (111) peak remained at about 2% of its initial value with pressure cycling. The insignificant change in the (111) peak width also indicates that domain size does not change under pressure (*14*, *23*). From this analysis, we determine that the increased peak width after unloading is caused by the formation of crystalline defects such as dislocations rather than changes in the size and shape of crystalline domains. The observation that (200) and (220) peak were the most affected and the (111) peak is least affected indicates the presence of stacking faults, twinning and dislocations (Fig. S6).

These XRD results were corroborated by high-pressure optical absorbance spectroscopy. Au nanocrystals have a plasmonic resonance that is dependent upon nanocrystal size, shape and microstructure (*24*). Previous optical modeling showed that the plasmon peak wavelength is indicative of nanocrystal shape, while an irreversible decrease in the plasmon peak intensity is indicative of the formation of crystalline defects (*2*). The plasmon peak wavelength of the 3.9 nm Au nanocrystals increased by ~30 nm when pressure was increased to 7.5 GPa and recovered its initial value upon unloading (Fig. S7). The plasmon peak intensity showed an irreversible decrease after unloading. This data supports the conclusion that the irreversible increase in XRD peak width after pressure cycling is due to the formation of crystalline defects, rather than a change in the size and shape of crystalline domains within the nanocrystals.

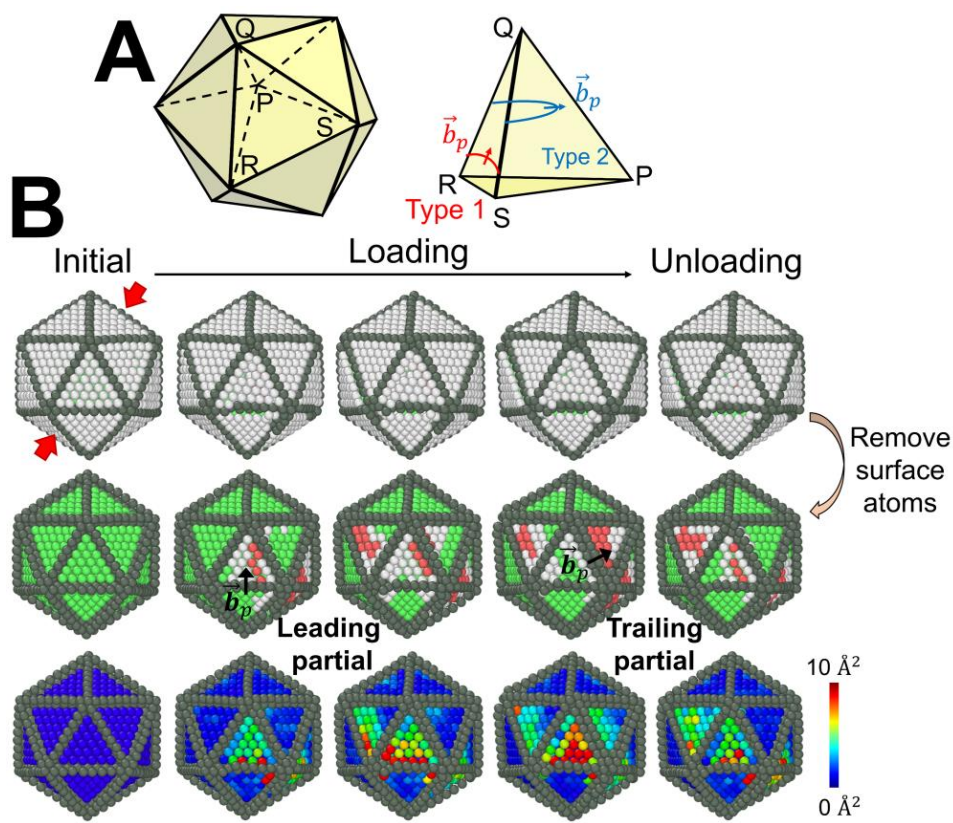

**Fig. 3. MD simulation of a 3.9 nm icosahedral nanocrystal**. A) Schematic of nanocrystal geometry and slip planes for stacking fault type 1 and type 2. B) Atomic configurations during loading and unloading process. Top row shows the surface atoms and the loading direction (red arrows). In the next two rows, outermost atoms are omitted to visualize the formation of defects. Images in middle row have green atoms for FCC, white atoms for unclassified crystal structure

(typically near the core of a partial dislocation or at the surface), and red atoms for HCP. Images in bottom row are colored according to non-affine squared displacement, in which the slip plane swept by a perfect dislocation is identified.

MD simulations were used to understand the crystalline defects that form within the nanocrystals, and their interactions with existing twin boundaries and surfaces. Two types of stacking faults (SF) were formed in an icosahedral nanocrystal under pressure (Fig. 3A); SF type 1 refers to a stacking fault parallel to the outer surface of the nanocrystal (or parallel to surface steps formed during deformation), and SF type 2 is a stacking fault parallel to an internal twin boundary that intersects with two other twin boundaries. Both types of stacking fault were formed by the nucleation and propagation of a Shockley partial dislocation with a Burgers vector of $\frac{1}{6}\langle 112 \rangle a$. SF type 1 forms when a Shockley partial dislocation with Burgers vector parallel to the outer surface propagates on a slip plane parallel to the outer surface. This results in a displacement relative to adjacent grains that is about the magnitude of the Burgers vector (see supplementary materials). When trailing partials are activated on the same plane, the stacking fault is removed, which results in the formation of a larger displacement. The trailing partial slip in one grain sometimes triggers stacking fault formation in an adjacent grain. This occurs if the Burgers vector of the trailing partial dislocation (i.e. the slip direction) is aligned well with the Burgers vector of a leading partial dislocation (Fig. 3B). SF type 2 is a dislocation that has a Burgers vector parallel to an interior twin boundary. The passage of SF type 2 is blocked by intersecting twin boundaries and forms interfacial dislocations with a $\frac{1}{9}\langle 222 \rangle a$ Burgers vector. This type of stacking fault has also been observed in penta-twinned silver nanowire with >40 nm diameter (*25*). In contrast to the penta-twinned silver nanowires, the trailing partial does not follow the leading partial (or, the SF type 2) in the 3.9 nm nanocrystal because the image stress is very large due to the proximity to the free surface and opposes the motion of the trailing partial. For this reason, SF type 2 is harder to form, and the plastic deformation of the nanocrystal is dominated by the successive formation of SF type 1 defects. This is in contrast with work by Sun et al. on Ag nanocrystals where they reported liquid like deformation via surface diffusion; however, they had performed very high temperature MD simulations to observe diffusion activity in MD time scale (*1*). We conducted room-temperature MD simulations where surface diffusion was limited. This is in line with experiments where the Au nanocrystal surface was protected by bulky organic ligands that form Au-SR bonds which prevent diffusion at the nanocrystal surface (*26*).

We attribute the irreversible deformation in the nanocrystals to SF type 1 defects, as portions of these defects remain in the simulated nanocrystal after unloading (See Fig. 3B). The stacking fault parallel to the outer surface is energetically meta-stable, because of the finite energy barrier required to form a partial dislocation to reversely sweep out the stacking fault. In experimental time scales, some meta-stable stacking faults can be expected to remain. In contrast, SF type 2 escapes quickly to the free surface during unloading upon the removal of deviatoric stress, which implies that the plastic deformation by this type of stacking fault is reversible. SF type 2 forms a partial dislocation loop that is blocked by twin boundaries. This is an unstable structure that is easily pulled towards the free surface by an image stress (*25*).

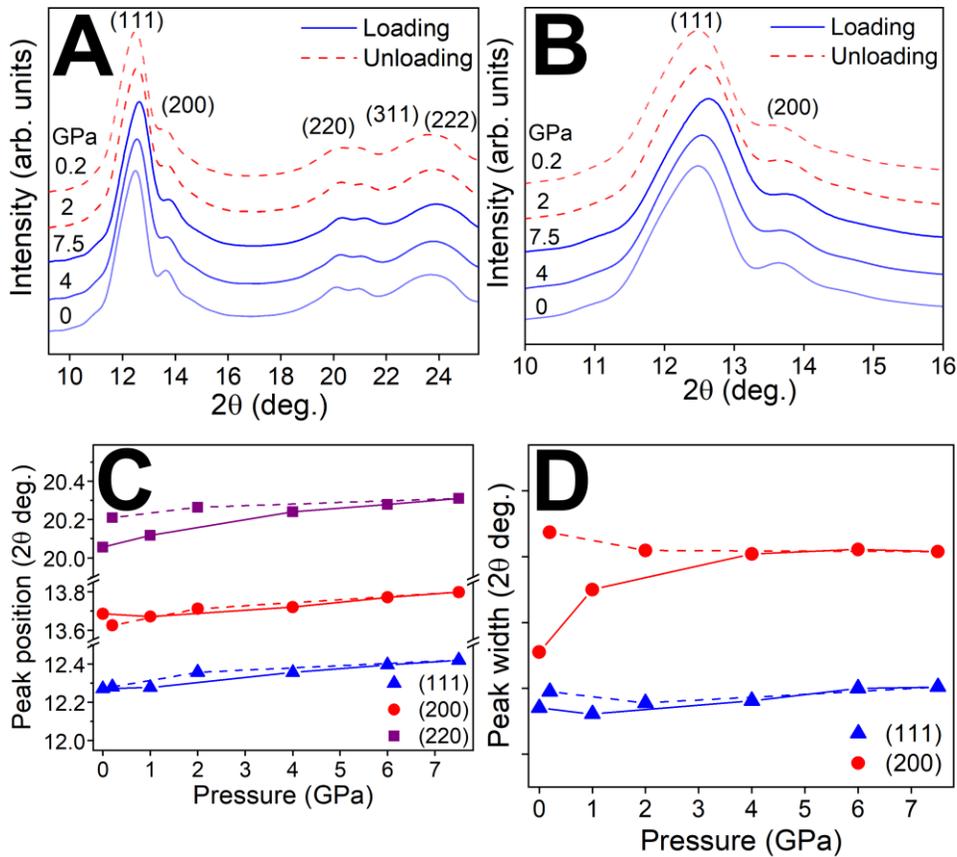

**Fig. 4. Simulated high-pressure XRD patterns from MD simulations.** A) All diffraction peaks and B) magnified view of (111) and (200) peaks. Change in diffraction peak C) position and D) width (each division is 0.1°), upon loading (solid line) and unloading (dashed line).

The correspondence between the experimental data and MD simulation was evaluated by generating XRD patterns from the MD simulated structures at different pressures by using the Debye scattering equation (*15*) (see Fig. 4A-B). The XRD peak width for the simulated patterns showed a similar trend to experimental data that the (111) peak width was least affect, and the (200) and (220) peaks broadened significantly under pressure (Fig. 4C-D). The effect of having more stacking faults in the nanocrystal is evident from the significant increase of peak width for the (200) and (220) peaks. XRD patterns from MD simulations cannot be fitted directly to the experimental high-pressure patterns as a complete size distribution was not simulated but, the qualitative effect of dislocation activity on XRD patterns is visible. The close agreement of XRD simulated from MD and experimental XRD patterns substantiate that MD simulations were a true representation of experiments.

In summary, using high-pressure XRD, optical absorbance spectroscopy and MD simulations we provide the first evidence of plastic deformation in individual 3.9 nm Au nanocrystals. The plastic deformation governed was by stacking faults formed via surface nucleated partial dislocations. The formation of surface steps during the passage of sequential partial dislocations as well as remaining stacking faults led to residual defects in the nanocrystal. The kinetics of residual defect recovery after unloading the sample will be explored further in future studies. This work provides a critical advancement in using experimental and simulation generated XRD as a comprehensive measurement technique to study defect formation in nanomaterials.

**Materials and Methods**

**Nanocrystal synthesis and characterization**

3.9 nm Au nanocrystals were synthesized according to Peng et al. (*12*). 20 ml tetralin was combined with 24.3 ml 70% oleylamine and 200 mg $HAuCl_4$ in air at $30^oC$. A reducing solution of 1 mmol tert-butylamine-borane complex, 2 ml tetralin and 2.4 ml 70% oleylamine was then rapidly injected into the solution. The reaction was allowed to proceed for 1 hour. Nanocrystals were precipitated using acetone, and collected through centrifugation at 8500 rpm for 3 min. Nanocrystals were redispersed in 100 ml toluene, and refluxed with 1 ml dodecanethiol for 15 min under $N_2$ gas. Nanocrystals were precipitated and washed using ethanol and redispersed in hexane. Nanocrystals were imaged using a FEI Technai G2 TEM at 200 keV accelerating voltage. Nanocrystal size distribution was determined from TEM images using ImageJ (Fig. S1).

**High pressure XRD**

Pressure-dependent measurements were performed in Diacell$^©$ One20 DAC from Almax easyLab with ruby powder as a pressure calibrant. The diamonds had 500 µm culets and a T-301 stainless steel gasket with a 300 µm hole was used. Nanocrystals were drop casted on a glass slide to form thick layer of gold superlattice. A small piece of the dried sample was loaded into the sample chamber with ruby powder and then the sample chamber was flooded with toluene. Toluene freezes at approximately 1.03 GPa and acts as a non-hydrostatic pressure medium (*27*). The mean stress was calculated from the shift in the R1 line (*28*).

XRD measurements were performed at beamline 12.2.2 at the Advanced Light Source at Lawrence Berkeley National Laboratory. The wavelength of the incident x-ray beam was fixed at 0.4974 Å and an x-ray spot size of 15 µm was used. Diffraction patterns were collected for 120 s using the Mar345 image plate detector. The sample to detector distance was calibrated using a CeO$_2$ standard. The 2D images were integrated to 1D plots using DIOPTAS software (*29*). The XRD peak parameters were calculated by fitting the peaks to a combination of Gaussian and Lorentzian peak functions along with a high order polynomial for the background.

## MD simulations

We carry out a series of MD simulations with LAMMPS software to investigate the mechanism of plastic deformation of Au nanoparticles (*30*). The amorphous phase matrix is modelled by a fictitious material where the atoms are interacted with the LJ potential $\left(\sigma = 2.56\,\text{Å},\ \varepsilon = 0.1\,\text{eV}\right)$ and its amorphous phase is constructed by an annealing-and-quenching process where we begin from the melted phase at 2000 K and slowly cool down it down to 300 K with the cooling rate of 34 K/ps. We did not consider the full atom modeling of the organic molecules in real amorphous matrix, because it costs very large computational time to achieve equilibrium phase composed of long molecules while the deformation mechanism of gold particle is not expected to be sensitive to the specific atom or molecules in the amorphous phase matrix. The interaction between the gold atoms is described by using EAM potential developed by H. Sheng (*31*) and the aforementioned LJ parameters is also used to describe the interaction between atoms in the matrix and the gold atoms. The gold nanoparticle is initially located at the center of the matrix, and equilibrated at 300K with NPT ensemble for 50 ps (*32*). We then applied compressive loading by moving the rigid plates at the top and bottom with the strain rate of 0.1/ns (e.g. $10^8$ /s). Periodic boundary conditions are applied in the transverse directions ($x, y$ directions) and a virial stress is used to compute the stress distribution of the system. We use the open visualization tool (OVITO) to visualize the atomic configurations, and employ the dislocation extraction algorithm (DXA) to identify dislocations and stacking faults (*33*).

## XRD from MD simulations

XRD patterns were generated from MD simulations for the 3.5 nm, 3.9 nm and 4.5 nm icosahedral nanocrystals and the 3.9 nm decahedral nanocrystal using the Debye scattering equation at different pressures (*14*). The simulated XRD patterns were calculated by fitting the peaks to a combination of Gaussian and Lorentzian functions along with a high order polynomial for the background, as with the experimental nanocrystals. The nanocrystal shape and size distribution from experiment was used to average the XRD pattern of individual simulated nanocrystals to generate the XRD pattern shown in Fig. 4, which represents the XRD pattern of an ensemble of simulated nanocrystals.

**High pressure optical absorbance spectroscopy**

Optical absorbance measurements were performed in a Bx90 DAC (DESY Deutsches Elektronen-Synchrotron, Hamburg, Germany) with type 2A diamonds (Technodiamant) with 300 μm culets (*34*). Stainless steel gaskets were indented to 50 μm and 125 μm holes were drilled by electric discharge milling. Ruby powder (Almax easyLab) was used as a pressure calibrant. Ruby fluorescence spectra were collected at each pressure on a confocal microscope (HORIBA) with 532 nm laser excitation. The mean stress was calculated from the shift in the R2 line.

Absorbance spectra of nanocrystals suspended in toluene as a non-hydrostatic pressure medium were collected a tungsten-halogen lamp mounted on an inverted Leica Dmi8 microscope. Transmitted light was collected with a 10x objective (Mitutoyo, NA=0.28) and a fiber-coupled Ocean Optics spectrometer. At each pressure, the lamp spectrum with and without the diamond anvil cell in the light path was measured in order to calculate the absorbance spectrum of the particles. The absorbance spectrum of the diamond anvil cell with only ruby powder was subtracted from the spectrum acquired at each pressure. For visualization purposes, spectra were processed with a low-pass filter to remove high-frequency noise.

**Acknowledgements:**

We thank Zhongwu Wang at Cornell High Energy Synchrotron Source for supporting this project. **Funding:** X.W.G. and A.P. acknowledge financial support from Stanford start-up funds. The Advanced Light Source is supported by the Director, Office of Science, Office of Basic Energy Sciences, of the U.S. Department of Energy under Contract No. DE-AC02-05CH11231. Beamline 12.2.2 is partially supported by COMPRES, the Consortium for Materials Properties Research in Earth Sciences under NSF Cooperative Agreement EAR 1606856. Part of this work was performed at the Stanford Nano Shared Facilities (SNSF), supported by the National Science Foundation under award ECCS-1542152. M.T.K. is supported by the National Defense and Science Engineering Graduate Fellowship. D.D. is supported by the NSF Graduate Fellowship. S.L. and S.R. are supported by the Creative Materials Discovery Program (2016M3D1A1900038) through the National Research Foundation of Korea (NRF) funded by the Ministry of Science and ICT. L.A.H. and K.A.H. acknowledge financial support from Trinity College. **Author contributions:** X.W.G. conceived the idea and supervised the research of this work. A.P. synthesized the particles and M.T.K performed the TEM characterization. A.P., M.T.K., D.D., M.K. and A.D. performed the high-pressure XRD. A.P. performed the XRD simulation and analysis. S.L. and S.R. performed the MD simulations and analysis. K.A.H. and L.A.H. performed the optical studies and analysis. A.P., S.L., L.A.H., S.R. and X.W.G. wrote the manuscript, and all the authors reviewed the manuscript.

Supplementary Material for

**Nucleation of Dislocations in 3.9 nm Nanocrystals at High Pressure**

Abhinav Parakh, Sangryun Lee, K. Anika Harkins, Mehrdad T. Kiani, David Doan, Martin Kunz,

Andrew Doran, Lindsey A. Hanson, Seunghwa Ryu, and X. Wendy Gu*

*To whom correspondence should be addressed. E-mail: xwgu@stanford.edu

**This PDF file includes:**

Supplementary Text

Fig. S1. TEM size distribution of as-synthesized and post-compression nanocrystals.

Fig. S2. XRD patterns of 3.9 nm Au nanocrystal at ambient pressure and generated bulk Au (ICDD PDF:00-004-0784).

Fig. S3. Rietveld-like refinement of ambient pressure XRD pattern using the Debye scattering equation.

Fig. S4. Changes in volumetric strain with pressure upon loading.

Fig. S5. Pressure dependence of calculated deviatoric stress for experiments and molecular dynamics (MD) simulation generated XRD patterns.

Fig. S6. Atomic configuration at the onset of plastic deformation showing the displacement jump.

Fig. S7. XRD patterns from various defects.

Fig. S8. Experimental high-pressure optical absorbance spectroscopy.

**Supplementary Text**

**Ambient pressure XRD**

Ambient pressure XRD for Au nanocrystals and bulk Au (from ICDD PDF: 00-004-0784) is plotted in Fig. S2.

**Debye scattering equation**

The Debye scattering equation was used to fit the ambient pressure XRD pattern to determine the structure of the nanocrystals. In this method, the atomic positions for 1 to 6 nm icosahedral and decahedral nanocrystals were generated. DebUsSy 2.0 software was used to simulate XRD patterns from the nanocrystals (*15*). A Rietveld-like refinement was done by controlling log-normal size distribution (2 variables), size dependent lattice strain function (4 variables), Debye-Waller factor and site occupancy (2 variables) for each particle shape and a distribution amongst the different nanocrystal shapes to fit the ambient pressure XRD pattern (a total of 16 variables) (*15*). This method for refinement resulted in the best fitting of the XRD patterns (Rw = 1.4%, see Fig. S3).

**Bulk modulus calculation**

The unit cell volume was obtained at different pressures by fitting the (111) and (220) diffraction peak. The modulus that corresponds to the change in volume versus pressure was found to be 226 GPa (Fig. S4). This is significantly higher than the bulk modulus for bulk Au (~170 GPa), and the previously reported bulk modulus for Au nanocrystals with sizes from 10 to 20 nm (~196 GPa) (*35*, *36*). The high value of the calculated modulus confirms the non-hydrostatic stress state within the diamond anvil cell and may have contributions from elastic size effects as well.

**Calculation of deviatoric stress**

The difference between the axial and radial stress (t) is calculated using lattice strain theory at each pressure (*37*). First, the quantity $Q(hkl)$ is calculated for the (111) and (220) peaks:

$$Q(hkl) = \frac{[a_m(hkl) - a_p]}{a_p(1 - 3\sin^2\theta_{hkl})} \tag{1}$$

Where $a_m$(hkl) is the lattice parameter from the experimental data (non-hydrostatic pressure), $a_p$ is the expected lattice parameter of Au under hydrostatic pressure, and $\theta_{hkl}$ is the experimental XRD peak position. $a_p$ is calculated by using 196 GPa as the bulk modulus of Au nanocrystals (*36*), and an effective hydrostatic pressure that is the sum of the applied pressure (measured from Ruby peak shift) and the pressure due to surface stress which was determined by the ambient pressure peak shift compared to bulk Au (Fig. S3). t is then calculated as:

$$t = (6G)\langle Q(hkl)\rangle f(x) \tag{2}$$

<Q(hkl)> is the average of Q(111) and Q(220). G is the shear modulus at the hydrostatic pressure (*38*). f(x) is equal to:

$$f(x) = \frac{A}{B} \tag{3}$$

Where A and B are constants that are defined as:

$$A = \frac{2x+3}{10} + \frac{5x}{2(3x+2)} \tag{4}$$

$$B = \alpha[x - 3(x-1)\langle\Gamma(hkl)\rangle] + \frac{5x(1-\alpha)}{3x+2} \tag{5}$$

$$x = \frac{2(S_{11} - S_{12})}{S_{44}} \tag{6}$$

$$\Gamma(hkl) = \frac{(h^2k^2 + k^2l^2 + l^2h^2)}{(h^2+k^2+l^2)^2} \tag{7}$$

α is equal to 0.5 (in between Reuss (iso-stress) and Voigt (iso-strain) conditions) (*21*). Γ(hkl) is calculated for the (111) and (220) peaks and then averaged to find <Γ(hkl)>. t as a function of applied pressure is shown in Fig. S5.

**XRD patterns from various defects**

As depicted in Fig. S7, we constructed atomic configurations of Au nanocrystals involving various defects, and obtained corresponding powder XRD patterns to better visualize the effect on XRD peak positions and widths. (200) and (220) XRD peaks are mostly affected by the stacking faults, while the effect of surface step is not significant.

**Optical absorbance spectroscopy**

The absorbance spectrum of the nanocrystals at ambient pressure shows a peak at 505 nm due to the plasmon resonance of the particles. With increasing pressure in the non-hydrostatic pressure medium, the plasmon red-shifts and increases in intensity. This observation is consistent with deformation of the quasi-spherical particles to ellipsoids in response to the uniaxial component of the stress applied. In addition, the absorbance efficiency increases upon pressurization followed by an irreversible decrease upon depressurization. The absorbance spectrum at ambient pressure after the pressurization cycle exhibits the same position and shape, but a decrease in intensity. These results are consistent with complete shape recovery accompanied by introduction of crystalline defects, as reported previously with compression to higher mean stresses (*2*).

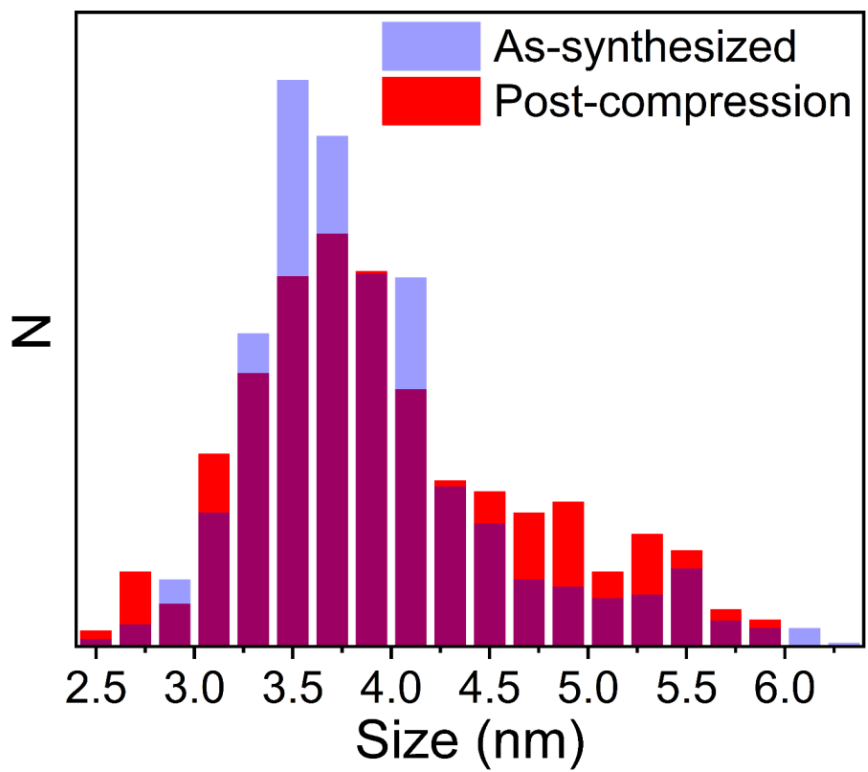

**Fig. S1. TEM size distribution of as-synthesized and post-compression nanocrystals.** The average diameter of as-synthesized nanocrystals was 3.93±0.65 nm (~800 particles measured) and of post-compression was 3.95±0.77 nm (~500 particles measured).

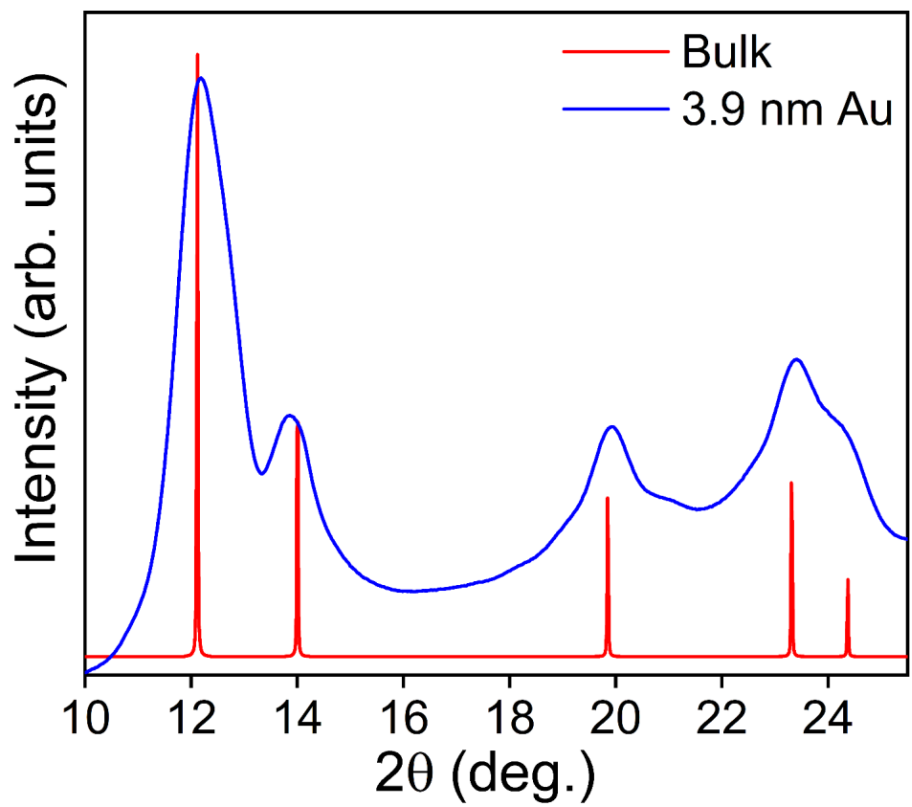

**Fig. S2. XRD patterns of 3.9 nm Au nanocrystal at ambient pressure and generated bulk Au (ICDD PDF:00-004-0784).**

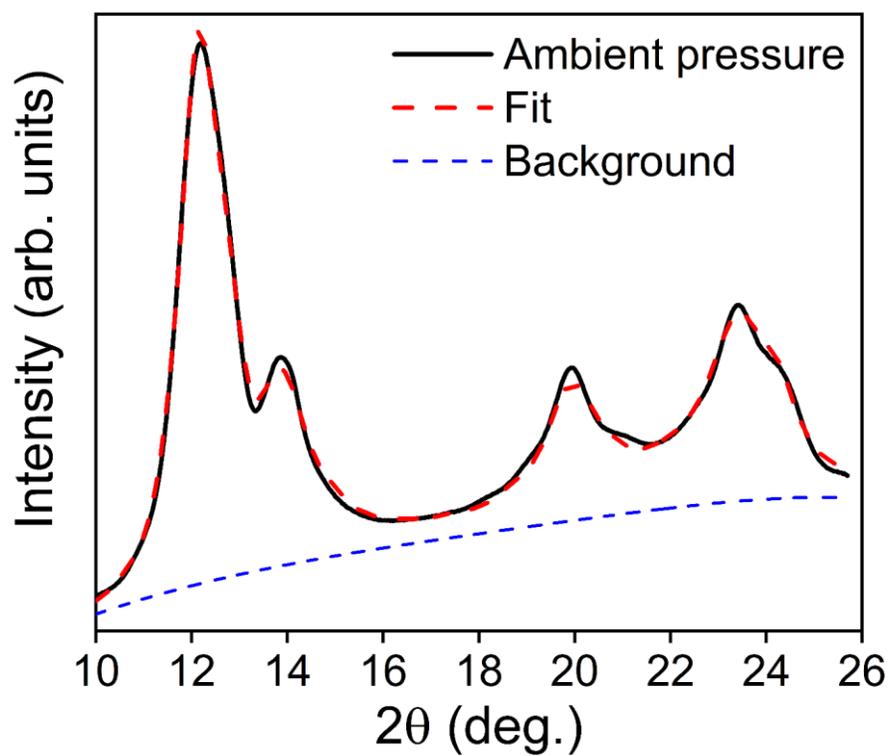

**Fig. S3. Rietveld-like refinement of ambient pressure XRD pattern using the Debye scattering equation.**

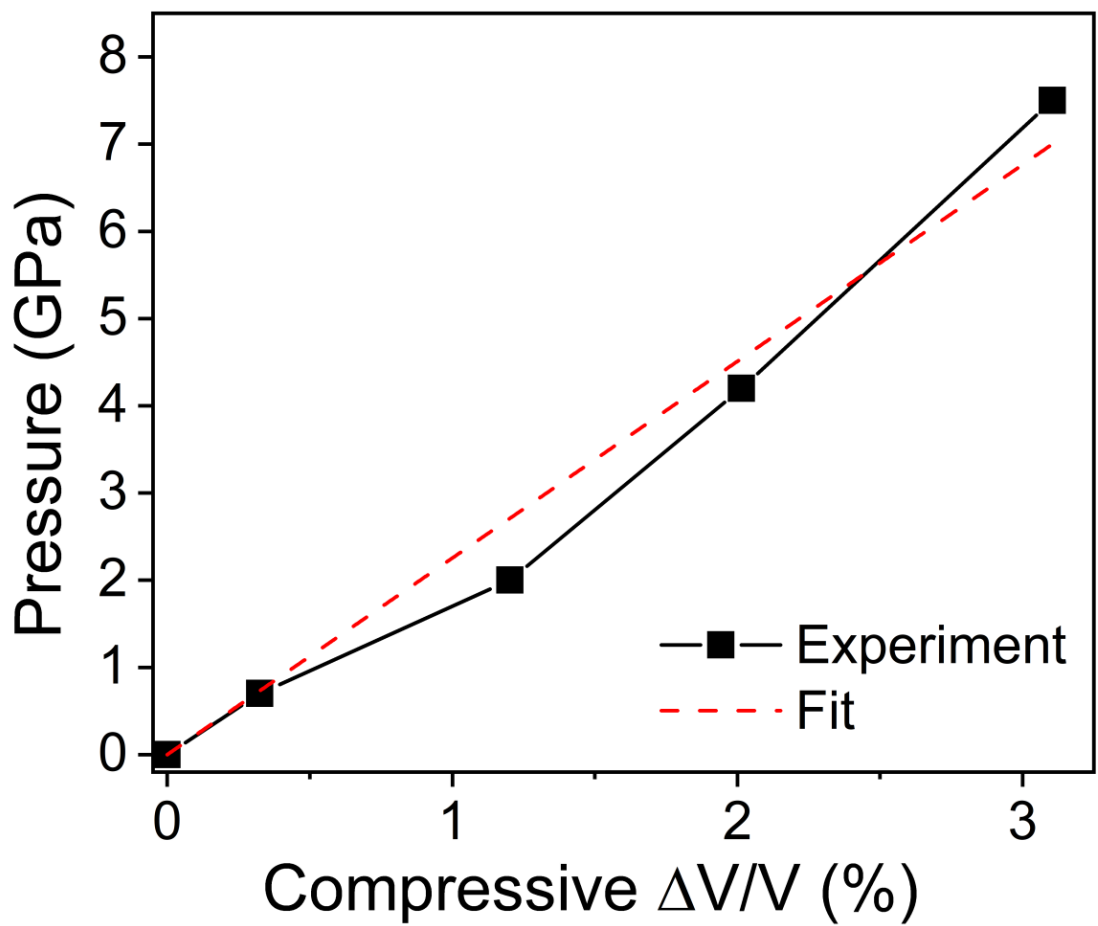

**Fig. S4. Changes in volumetric strain with pressure upon loading.** Linear fit was used to determine the bulk modulus for the Au nanocrystals.

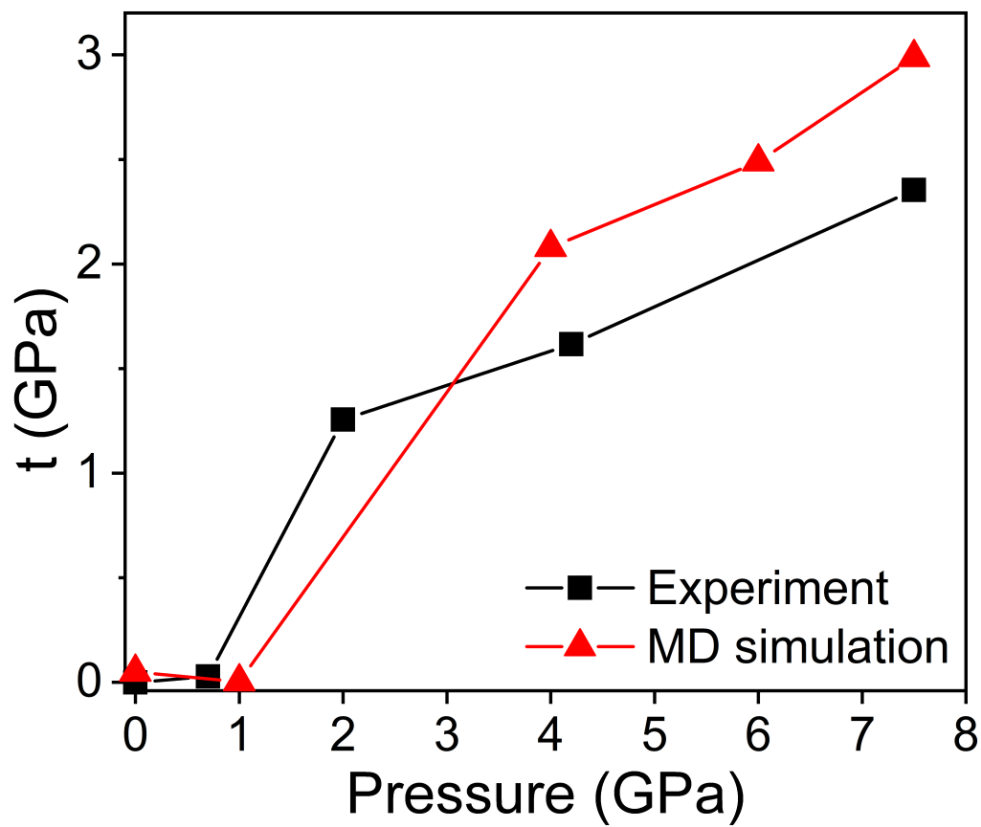

**Fig. S5. Pressure dependence of calculated deviatoric stress for experiments and molecular dynamics (MD) simulation generated XRD patterns.**

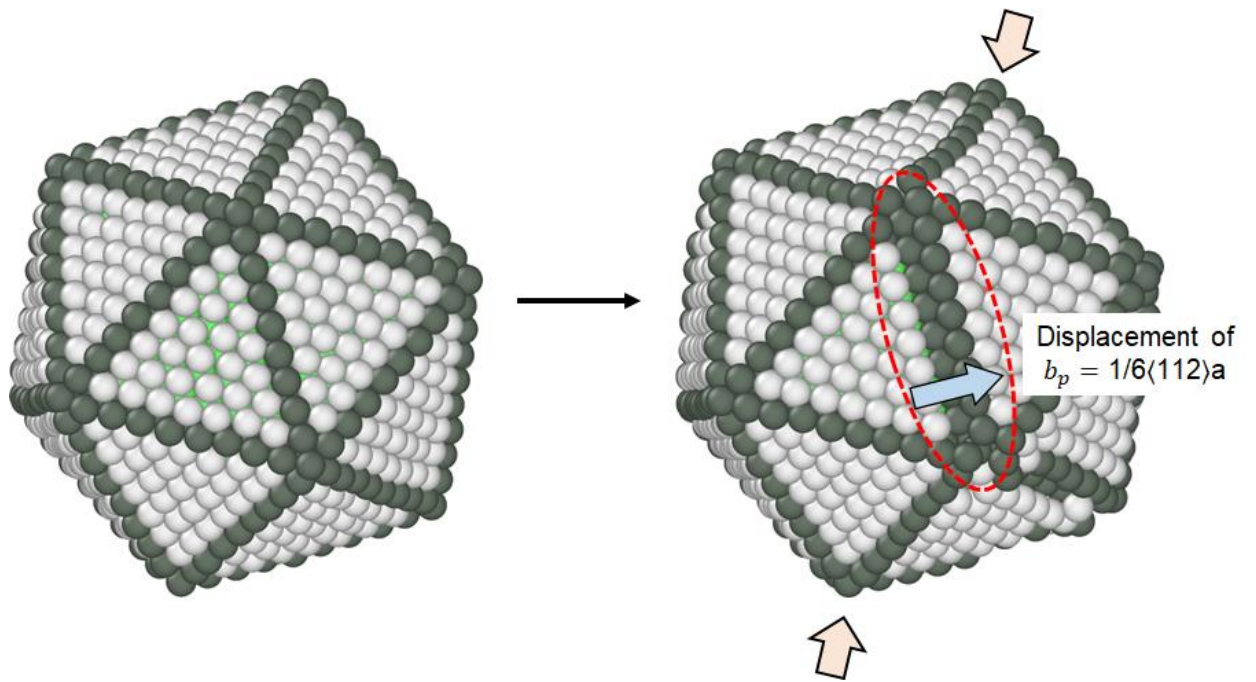

**Fig. S6. Atomic configuration at the onset of plastic deformation showing the displacement jump.**

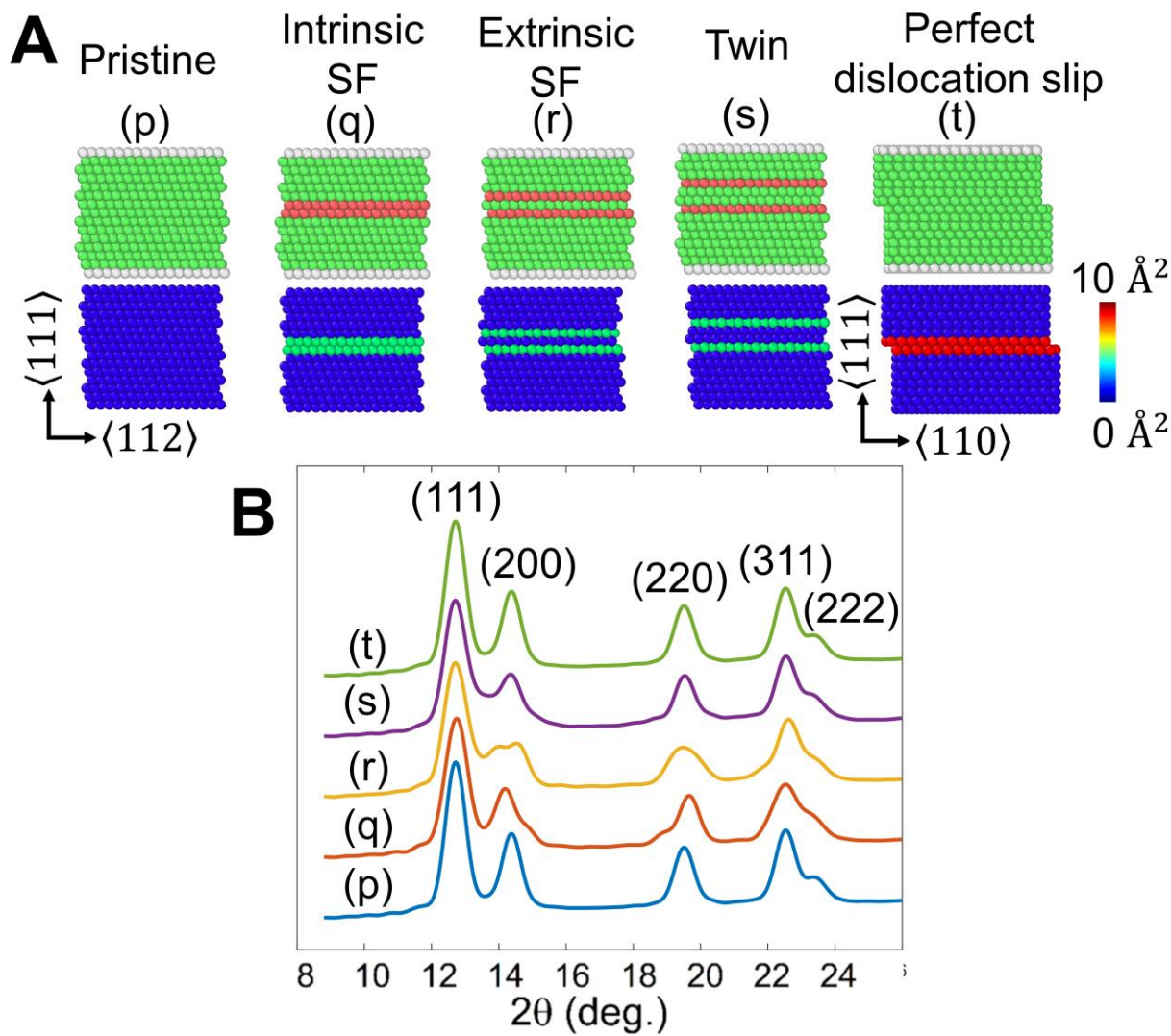

**Fig. S7. XRD patterns from various defects.** A) Atomic configurations of Au cubic nanoparticles involving (p) no defect, (q) intrinsic stacking fault (SF), (r) extrinsic SF, (s) twin, (t) surface step after perfect dislocation sweeping. Upper rows are visualized according to centro-symmetry parameters; green for FCC atoms, white for surface atoms, and red for HCP atoms. Lower rows are visualized according to non-affine squared displacement which is able to identify the slip plane swept by a perfect dislocation. B) Powder XRD patterns from five different Au cubic nanoparticles.

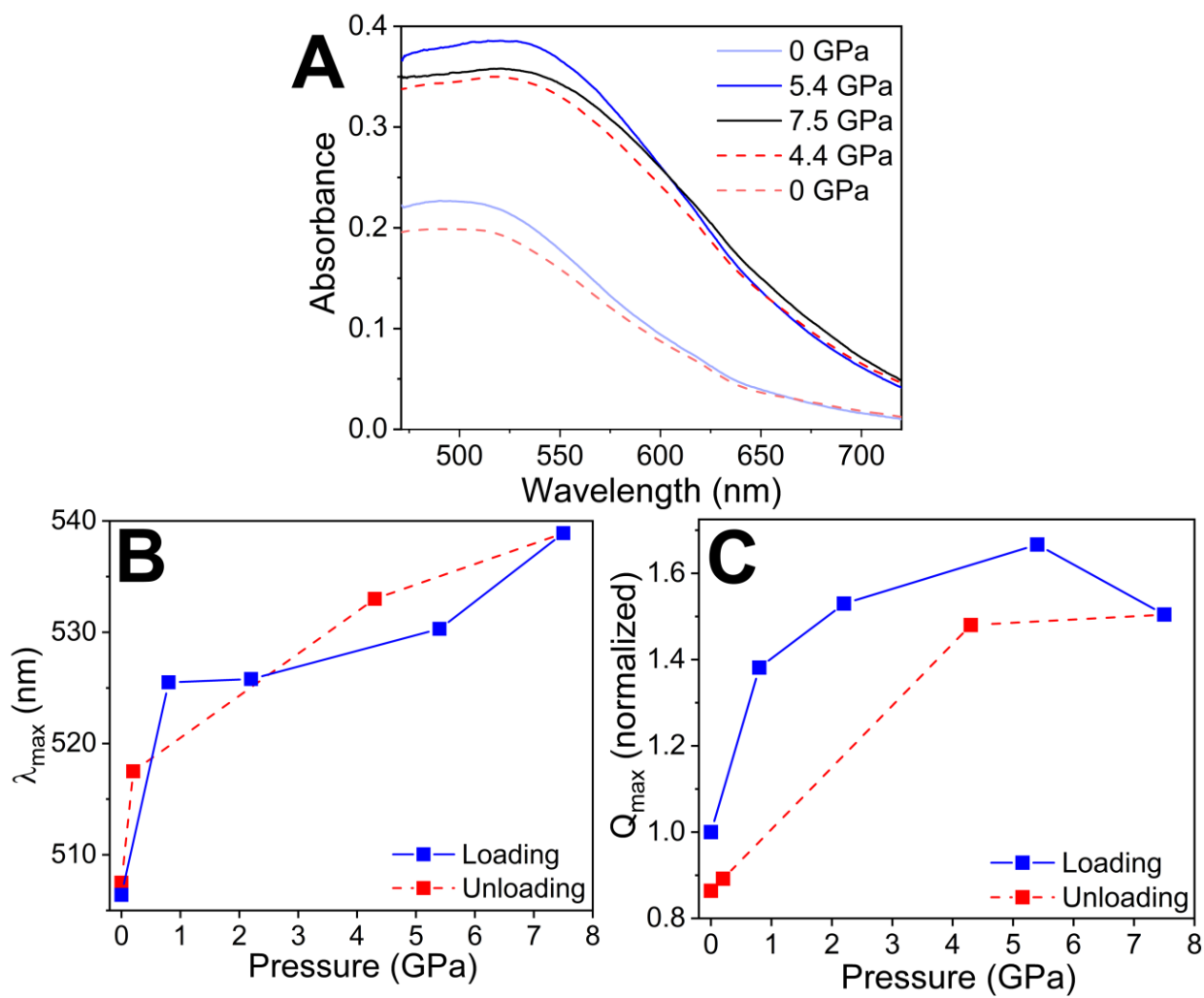

**Fig. S8. Experimental high-pressure optical absorbance spectroscopy.** A) Absorbance spectra of 3.9 nm Au nanocrystals in toluene at varied pressure. Spectra acquired while increasing pressure are shown in solid lines, and spectra acquired while decreasing pressure are shown in dashed lines. B) Peak wavelength of the plasmon resonance versus pressure. The peak position recovers to its initial position upon depressurization. C) Absorbance efficiency versus pressure. The absorbance efficiency shows an irreversible decrease after pressurization to 7.5 GPa.